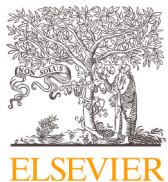
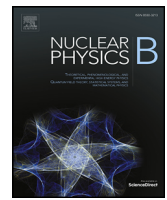

# Thermodynamic analysis of a compact object in Rastall–Rainbow gravity

Sareh Eslamzadeh, Saheb Soroushfar  *

*Department of Physics, College of Sciences, Yasouj University, 75918-74934, Yasouj, Iran*

A B S T R A C T

In this paper, we investigate the thermodynamic behavior of a horizonless compact object within the framework of Rastall–Rainbow (RR) gravity. Working with local shell thermodynamics for gravastar and an exterior fiducial temperature, we show that the RR modification bends temperature to produce two extrema and a stable mass remnant at zero temperature. We show that the gravastar's shell entropy is smaller than that of a comparable black hole, and that RR modifications introduce a logarithmic correction which contributes to specific heat positivity and a smoother free energy landscape of small gravastars.

A Central finding of this work is that, from heat capacity and Helmholtz free energy analyses, we uncover small, middle, and large branches and demonstrate that unlike Rainbow modified black holes, the small RR gravastar is both locally and globally favored over hot curved space. At the parameter level, both Rastall and Rainbow play distinct roles. Increasing the Rastall parameter, by strengthening matter–curvature coupling, adjusts the redshift between the shell and the exterior, shifts the temperature maximum to higher values at larger masses, and narrows the unstable window. In contrast, increasing the Rainbow parameter enhances energy dependent UV suppression and bends the temperature in lower values at larger masses. Altogether, these results highlight a controlled route to thermodynamic stabilization and the emergence of a stable remnant in horizonless compact objects within RR gravity.

## 1. Introduction

Despite extensive observational and theoretical progress since Einstein formulated general relativity, the need to explore alternative theories remains clear. It is essential to redefine some fundamentals, both at large scales where matter interacts with the underlying geometry of spacetime, and at short scales and high energies where quantum theory meets general relativity. Among the options available, we have chosen two intriguing alternative theories, Rastall and Rainbow, as well as their combination.

How does the accelerated universe, containing dark energy, affect the structures? More fundamentally, we can reframe the question to: how is the coupling between curvature and matter determined? We know from the Einstein field equations that the energy–momentum tensor is covariantly conserved, i.e., $T^{\mu\nu}_{;\mu} = 0$. Among enormous theories that get involved in this issue, Rastall's theory [1] challenges an assumption of the conservation law for the energy–momentum tensor by permitting its non-conservation, allowing for certain dark energy components to couple non-minimally with curvature. This suggests that adding new terms to Einstein's field equations is allowed in highly curved spacetime. Here, the standard conservation law is modified to $T^{\mu\nu}_{;\mu} = \lambda R^{,\nu}$, where $T$ and $R$ are

---

an energy–momentum tensor and Ricci scalar, respectively, while $\lambda$ is a constant. In recent years, this theory has garnered significant interest, particularly regarding its *cosmological implications*. Notably, several studies in Refs. [2–8] have investigated Rastall's cosmological features. For example, authors in Ref. [2] examined the consistency of Rastall's modification with the observational data in a matter-dominated FLRW universe. This agreement was further verified in Ref. [3] using SNe Ia and GRB data, though the model predicted a larger mass-density parameter than cluster gas mass fraction data, as it accounted only for matter without considering particle production due to non-conservation of the stress-energy tensor. Additionally, Refs. [4,5] introduced Rastall cosmology with two field components, and the Big Bang singularity problem was explored in Ref. [6]. The authors of Ref. [8] studied the coupling of radiation fields with geometry during the inflation era, and a recent work compared Rastall-$\Lambda$CDM cosmologies using various databases. Subsequent work examined *compact objects* in Rastall gravity. For instance, research has investigated neutron stars [9], static, spherically symmetric solutions [10], black holes surrounded by a perfect fluid [11], Quintessence compact stars [12], and gravastars [13] in Rastall gravity. In Ref. [10], exact solutions were studied, focusing on the role of Rastall parameter, cases where solutions resemble those in k-essence theories and the impact of scalar field potentials on solution stability. In Ref. [11], Kiselev–like black holes have been examined in the context of Rastall theory. In particular, authors in Ref. [13] have provided comprehensive research on Rastall's effects for three regions of gravastars. Additionally, the thermodynamics of black holes [14,15] and the geodesic equations of black holes surrounded by perfect fluid [16] have also been explored within Rastall theory and more recently a neutral regular black hole with generalized uncertainty principle (GUP) corrections has been studied in this framework [17].

Beyond such a large-scale modification lies the challenge of unifying quantum theory and gravity. As a matter of fact, the Planck scale plays a fundamental role in various proposals, such as string theory, loop quantum gravity, noncommutative geometry. Motivated by this, Double Special Relativity (DSR) [18,19] was developed, treating the Planck energy as a universal constant; within this framework, Magueijo and Smolin introduced Rainbow gravity [20], which departs from classical gravity in two key respects. First, regarding the modified special relativity, a free-falling particle with energy $E$ observes the same physics laws; Second, in Rainbow gravity, spacetime depends on the observer's energy, a feature absent in classical descriptions of gravity. Numerous studies have examined the thermodynamical aspects of black holes in Rainbow gravity. For instance, Ref. [21] focused on temperature and entropy, Ref. [22] analyzed Hawking radiation via tunneling, Ref. [23] investigated phase transitions, and Ref. [24] explored geometrical thermodynamics. In addition, Ref. [25] presented a comprehensive study of black hole remnants, including Schwarzschild, Kerr, Kerr-Newman, and charged AdS black holes. Beyond thermodynamics, further studies in the framework of Rainbow gravity have addressed the dynamical and structural properties of black holes. Among them, Ref. [26] studied the geodesic structure of a Schwarzschild black hole, Ref. [27] compared the in-fall time for different observers, and Ref. [28] analyzed the energy required for information duplication by calculating the critical Rainbow parameter. Most recently, Ref. [29] investigated the structural properties of dark energy stars in Rainbow gravity.

As a synthesis of the above theories, Rastall–Rainbow (RR) gravity has emerged as an innovative approach. Mota et al. have extensively explored this theory in Ref. [30], where they first modified Einstein's equations according to Rastall's theory and then adjusted them to construct an energy–dependent metric function that satisfies Rainbow theory, as we review in Section 2. Along with this new approach, the authors continued their research in Refs. [30,31] by deriving the modified pressure-radius and mass-radius relations, known as the Tolman–Oppenheimer–Volkoff (TOV) equations [32,33], for neutron stars. They employed three different equations of state to model the nuclear pressure of neutron stars against gravitational collapse and determined the maximum mass of compact objects for various values of the Rainbow and Rastall parameters for each model. Additionally, they applied their findings to specific astrophysical objects. Furthermore, within the framework of RR gravity, charged gravastars [34], charged anisotropic condensed stars [35], white dwarfs [36], Bose–Einstein condensate stars [37], and, more recently, stable quark stars [38] have been investigated. Thus far, most RR studies of compact objects have remained at the level of modified TOV structure, tuning RR parameters to shift mass-radius relations or widen equilibrium windows, without offering a thermodynamic mechanism. To go beyond, we develop a thermodynamic framework for a horizonless compact object in RR gravity and use it to analyze the gravastar remnant, stability, and phase transitions. This shifts the focus from structure to thermodynamics and clarifies the distinct roles of the Rainbow and Rastall modifications in the thermal behavior.

We aim to study the thermodynamic features of a compact object in RR gravity. Our research is organized as follows. In Section 2, we provide a review of RR gravity as the basis for our calculations. In Section 3, we explore the compact object's temperature, heat capacity, free energy, and phase transitions in RR gravity. Finally, we summarize our results in Section 4.

## 2. Rastall–Rainbow theory

At extremely small length scales (near the Planck length) or equivalently at extremely high energies (close to the Planck energy), the interplay between gravity, quantum mechanics, and special relativity becomes non-negligible. This raises a fundamental question: in which inertial frames are these thresholds meaningful? There are two primary approaches to address this issue. One is to accept Lorentz symmetry breaking at the Planck scale, implying the existence of a preferred reference frame. The alternative is to preserve Lorentz invariance by modifying the Lorentz transformations in momentum space. This second viewpoint led to the development of DSR [18,19], which introduces a second invariant scale, the Planck energy, in addition to the speed of light.

To extend DSR to curved spacetimes, Magueijo and Smolin proposed Rainbow gravity [20], where the background geometry depends on the energy of the test particle. In their formulation, the spacetime metric becomes energy-dependent, leading to a family of metrics parameterized by the ratio $\varepsilon = E/E_P$. In particular, the modified dispersion relation (MDR) is given in Ref. [20] as

$$\Pi^2(\varepsilon) E^2 - \Sigma^2(\varepsilon) p^2 = m^2, \tag{1}$$

where $\Pi(\varepsilon)$ and $\Sigma(\varepsilon)$ are Rainbow functions that modify the temporal and spatial components of the metric, respectively. Among the various proposed forms, we adopt a widely used one inspired by loop quantum gravity and non-commutative geometry (see, e.g., Refs. [25,28,39–42]), given by

$$\Pi(\varepsilon) = 1, \quad \Sigma(\varepsilon) = \sqrt{1 - \gamma \varepsilon^2}, \tag{2}$$

where $\gamma$ is a dimensionless parameter encoding the strength of quantum gravitational effects. This choice ensures that the standard energy–momentum relation is recovered in the low-energy limit ($\varepsilon \to 0$).

Accordingly, the energy–dependent line element for a static spherically symmetric spacetime becomes

$$ds^2 = -\frac{B(r)}{\Pi^2(\varepsilon)} dt^2 + \frac{A(r)}{\Sigma^2(\varepsilon)} dr^2 + \frac{r^2}{\Sigma^2(\varepsilon)} (d\theta^2 + \sin^2\theta \, d\phi^2). \tag{3}$$

In our case, since $\Pi(\varepsilon) = 1$, the speed of light is defined as

$$c(\varepsilon) = \frac{\Pi(\varepsilon)}{\Sigma(\varepsilon)} = \frac{1}{\sqrt{1 - \gamma \varepsilon^2}}, \tag{4}$$

which increases with energy and diverges as $\varepsilon \to \gamma^{-1/2}$, reflecting the existence of a maximal energy cutoff. This behavior is consistent with DSR predictions and suggests the emergence of a minimal length scale in nature.

On the other hand, the Rastall theory modifies the standard conservation law of the energy–momentum tensor. As originally proposed in Ref. [1], instead of $T^{\mu\nu}{}_{;\mu} = 0$, one adopts

$$T^{\mu\nu}{}_{;\mu} = \lambda \nabla^\nu R, \tag{5}$$

where $\lambda$ is the Rastall parameter and $R$ is the Ricci scalar. This leads to the modified Einstein field equations

$$R^\nu_\mu - \frac{\lambda}{2} \delta^\nu_\mu R = 8\pi G \, T^\nu_\mu. \tag{6}$$

Combining both Rainbow and Rastall modifications, the RR field equations become

$$R^\nu_\mu(\varepsilon) - \frac{\lambda}{2} \delta^\nu_\mu R(\varepsilon) = 8\pi G(\varepsilon) \, T^\nu_\mu(\varepsilon). \tag{7}$$

To describe matter, we use the energy–momentum tensor of a perfect fluid

$$T_{\mu\nu} = p(r) \, g_{\mu\nu} + [p(r) + \rho(r)] \, U_\mu U_\nu, \tag{8}$$

with the four-velocity given by

$$U^\mu = \left( \frac{1}{\sqrt{B(r)}}, 0, 0, 0 \right). \tag{9}$$

By solving the field equations in Eq. (7), we derive the metric function presented in Eq. (3) as follows

$$A(r) = \left[ 1 - \frac{2GM(r)}{r} \right]^{-1}, \tag{10}$$

where the mass function is defined as

$$M(r) = \int_0^R 4\pi r'^2 \, \bar{\rho}(r') \, dr'. \tag{11}$$

The other metric component in Eq. (3) can be obtained by solving the remaining field equations in Eq. (7) through the following relations

$$\frac{B'}{2B} = \frac{GM}{r^2} \left( 1 + \frac{4\pi r^3 \bar{p}}{M} \right) \left( 1 - \frac{2GM}{r} \right)^{-1}, \tag{12}$$

$$\frac{B'}{2B} = -\frac{\bar{p}'}{\bar{p} + \bar{\rho}}. \tag{13}$$

In the preceding equations, the effective energy density and pressure are introduced, taking into account the contributions from both the Rastall coupling and Rainbow modifications, as

$$\bar{\rho} = \frac{1}{\Sigma^2(\varepsilon)} [\alpha_1 \, \rho + 3\alpha_2 \, p], \tag{14}$$

$$\bar{p} = \frac{1}{\Sigma^2(\varepsilon)} [\alpha_2 \, \rho + (1 - 3\alpha_2) \, p], \tag{15}$$

with the coefficients $\alpha_i$ determined explicitly by the Rastall parameter as

$$\alpha_1 = \frac{1 - 3\lambda}{2(1 - 2\lambda)}, \quad \alpha_2 = \frac{1 - \lambda}{2(1 - 2\lambda)}.$$

This framework sets the stage for investigating thermodynamic properties of compact objects in RR gravity, as we develop in the following sections.

## 3. Thermodynamics and phase transition of a compact object in rastall–Rainbow theory

A compelling alternative to classical black holes is the gravitational vacuum condensate star, or "gravastar," originally proposed by Mazur and Mottola [43,44]. In this model, spacetime is divided into three distinct regions: (i) an interior de Sitter-like vacuum with the equation of state $p = -\rho$, (ii) a thin shell of ultra-stiff matter with $p = +\rho$, and (iii) an exterior vacuum described by the Schwarzschild solution. The repulsive pressure of the interior is balanced by the shell tension, which acts as a quantum phase boundary. Although lacking an event horizon or singularity, gravastars can closely mimic the exterior spacetime of black holes for distant observers.

This concept has motivated extensive theoretical and observational studies exploring gravastar properties from multiple perspectives. First, the thermodynamic aspects, including entropy, internal energy, and temperature profiles, have been addressed in both original and extended models [43]. Afterwards, investigations have examined their shadow formation [45], quasi-normal modes [46], gravitational lensing signatures [47], and radiation characteristics [48,49]. Subsequent developments have analyzed structural deformations through tidal Love numbers and multipole moments [50], the compatibility of gravastars with shadow observations such as M87* [51,52], and their potential role as non-singular alternatives to black holes in regularized spacetime models [53]. The effects of graviton mass on shell dynamics have been investigated within massive gravity frameworks [54], while extensions of teleparallel gravity have linked dark energy and symmetry properties to gravastar stability [55]. Collectively, these studies emphasize the continued relevance of gravastars as viable candidates for horizonless compact objects within both classical and modified gravity contexts.

Building on the combined RR framework outlined in the Introduction and the gravastar model summarized above, we investigate horizonless compact objects in the RR setting. While black holes in Rainbow and Rastall gravity have been extensively studied, the thermodynamic aspects of RR gravastars, apart from rare cases such as the charged RR gravastar in Ref. [34], remain under-explored, leaving significant scope for further study. In the next section, we focus on how RR deformations shape the internal geometry and thermodynamics of gravastars.

*3.1. Fiducial temperature and entropy*

**Setup & Assumptions:** We consider a horizonless compact object, namely a gravastar, not within standard GR but in the modified RR gravity. The framework features two modifications: a Rastall parameter encoding non-minimal matter-curvature coupling, and a Rainbow parameter introducing an energy-dependent metric that implements an effective UV cutoff motivated by quantum gravity scenarios. As noted earlier, the gravastar consists of three regions. Accordingly, in the shell we will compute the local temperature and entropy, while in the exterior region we will evaluate the fiducial temperature for an observer at a fixed radius, the heat capacity, and the Helmholtz free energy. The ultimate goal is to establish the thermodynamic behavior of a horizonless compact object in the RR background, with particular emphasis on its stability and phase transitions under variations of the modification parameters. In the following, we explore in three regions and analyze each separately within the framework of RR modification.

**A1. Inside**

Considering modified Einstein equations based on RR gravity, for the interior region, $0 \leqslant r \leq R$, with equation of state $p = -\rho$, Eqs. (12)–(15) lead to the following result

$$B^{-1}(r) = A(r) = \left[1 - \frac{2GM}{r}\right]^{-1}, \tag{16}$$

where $M$ is determined by Eq. (11) as follows

$$M = \left[\frac{\rho}{(2\lambda - 1)\,\Sigma^2(\epsilon)}\right] \frac{4\pi R^3}{3}. \tag{17}$$

Obviously, the mass function depends on Rastall parameter and Rainbow functions. Therefore, considering Eq. (3) we can obtain the metric function for the interior region in the following form

$$ds^2 = -\frac{1}{\Pi^2}\left(1 - \frac{2GM}{r(2\lambda-1)\Sigma^2(\epsilon)}\right)dt^2 + \frac{1}{\Sigma^2(\epsilon)}\left(1 - \frac{2GM}{r(2\lambda-1)\Sigma^2}\right)^{-1}dr^2 + \frac{r^2}{\Sigma^2(\epsilon)}\,d\Omega^2. \tag{18}$$

where $M$ is the total mass of the interior region.

**A2. Intermediate Thin Shell**

The thin shell, defined by the region $R - \delta < r < R + \delta$, is assumed to be filled with an ultra-relativistic fluid obeying the equation of state $p = \rho$. Several studies have explored the properties of such a shell, including its proper thickness, stability, and the associated junction and boundary conditions; see, e.g., Refs. [13,34]. For the present analysis, it suffices to determine the corresponding metric functions by solving Eq. (3), which yields

$$ds^2 = -\frac{1}{\Pi^2}\left(1 - \frac{2GM(2-3\lambda)}{r(1-2\lambda)\Sigma^2}\right)dt^2 + \frac{1}{\Sigma^2}\left(1 - \frac{2GM(2-3\lambda)}{r(1-2\lambda)\Sigma^2}\right)^{-1}dr^2 + \frac{r^2}{\Sigma^2}\,d\Omega^2. \tag{19}$$

*Local Temperature*

In the original gravastar model proposed by Mazur and Mottola [43,44], the local temperature of the shell is derived based on thermodynamic and quantum considerations, rather than from surface gravity. The key assumptions in this derivation are as follows:

- The shell is composed of an ultra-stiff quantum fluid satisfying the equation of state $p = \rho$.
- The chemical potential is neglected, $\mu = 0$, leading to the simplified Gibbs relation: $d\rho = T\,ds$.
- It is assumed that the energy density scales with the square of the local temperature, inspired by quantum field behavior in lower-dimensional Casimir–like systems

$$\rho = \left(\frac{a^2}{8\pi G}\right) T^2, \tag{20}$$

where $a$ is a dimensionless constant.

Solving for the temperature gives the local proper temperature of the shell as follows

$$T_{on-shell}^{RR} = \sqrt{\frac{8\pi G}{a^2}\rho}\,. \tag{21}$$

Here, the dimensionless constant $a$ is set to $a = 4\pi\sqrt{3}$, such that in the limiting case $\lambda \to 0$ and $\Sigma(\epsilon) \to 1$, the resulting expression exactly reproduces the Hawking temperature, $T = \frac{1}{8\pi GM}$, thereby preserving the same inverse mass dependence, even though the gravastar possesses no event horizon. Moreover, it is important to note that although the temperature is derived based on the shell's local thermodynamic properties, the mass $M$ appearing in the temperature relation refers to the interior mass of the gravastar (Eq. (17)). This is because, in the Mazur–Mottola model, the dominant contribution to the total gravitational mass arises from the interior energy density, while the shell mainly serves as a quantum phase boundary with negligible energy content. Therefore, the inverse mass dependence in the temperature scaling reflects the gravitational energy of the core rather than that of the shell. Thus, the Hawking–like temperature of the compact object on the intermediate thin shell can be expressed either as a function of the gravastar radius or as a function of its mass, as given by the following two equivalent equations

$$T_{on-shell}^{RR} = \frac{1}{4\pi R} \cdot (2\lambda - 1)\Sigma^2(\epsilon), \tag{22}$$

$$T_{on-shell}^{RR} = \frac{1}{8\pi GM} \cdot (2\lambda - 1)\left(1 - \frac{\gamma}{4\,GM^2}\right). \tag{23}$$

As seen from Eq. (22), the obtained temperature for the gravastar is clearly lower than the standard Hawking temperature. This reduced temperature implies a slower evaporation process and the presence of a stable remnant. Such behavior is a common feature in many quantum gravity-inspired models and may help avoid the final singularity and address the information loss paradox. To identify the existence of this remnant – corresponding to the point where the temperature vanishes at a finite mass, $M_0$, – we further reformulate the temperature expression in Eq. (23). To derive that, we first estimate the energy scale $\epsilon = E/E_P$ that enters the Rainbow functions. This is done by applying the Heisenberg Uncertainty Principle (HUP) in the form $E \sim 1/\Delta x$, where $\Delta x$ represents the characteristic length scale of the gravitating system. For simplicity and consistency with black hole thermodynamics, we approximate this length scale by the Schwarzschild radius, i.e., $\Delta x \sim 2GM$. Substituting this into the definition of $\epsilon$, and using $E_P = 1/\sqrt{G}$, we obtain the compact form $\epsilon = \frac{1}{2\sqrt{G}M}$, which is then used in the Rainbow-modified temperature relation. It is worth noting that, for black hole spacetimes, it is customary to set $\Delta x \sim r_{Sch} = 2GM$, corresponding to the Schwarzschild radius. Although the effective radius of the gravastar in the RR gravity framework is larger than the Schwarzschild radius by a factor of $\frac{1}{(2\lambda-1)\Sigma^2(\epsilon)}$, it follows that the corresponding remnant mass of the gravastar is smaller than that of a black hole, namely $M_0 = \frac{\gamma}{4G}$. Nevertheless, using the Schwarzschild radius as the characteristic length scale in the uncertainty relation remains a sufficiently accurate approximation for our analysis at this stage.

*Entropy of the Thin Shell*

In the gravastar framework adapted to RR gravity, entropy is localized entirely within the intermediate thin shell region, where the interior de Sitter core is treated as a condensate with zero entropy [44]. As we know, the shell consists of an ultra-stiff quantum fluid satisfying the equation of state and thermodynamic quantities are derived via local considerations rather than horizon-based definitions. Assuming a local Gibbs relation in differential form, $d\rho = T\,ds$, and considering the ultra-stiff fluid in the shell to exhibit a Casimir-like behavior with $\rho \propto T^2$, it follows that $s \propto T \propto \sqrt{\rho}$. Under the condition $p = \rho$, this leads to the local entropy density in the shell is modeled as

$$s(r) = a\sqrt{p(r)}, \tag{24}$$

where $a$ is a dimensionless constant fixed to reproduce the Hawking limit in the appropriate regime. Therefore, the total entropy in the shell, extending from $r_1 = R - \delta$ to $r_2 = R + \delta$, is given by [44]

$$S_{on-shell}^{RR} = 4\pi \int_{R-\delta}^{R+\delta} \frac{a}{\Sigma(\epsilon)}\sqrt{p(r)A(r)}\,r^2 dr. \tag{25}$$

Assuming a very thin shell ($\delta \ll R$), we apply the zeroth-order Taylor expansion to approximate the integrated around $r = R$, leading to the simplified expression

$$S_{on-shell}^{RR} \approx \frac{8\pi a\delta}{\Sigma(\epsilon)}\,R^2\sqrt{p(R)\,A(R)}. \tag{26}$$

This expression will be used in the following sections to compute the entropy and investigate the thermodynamic behavior of the gravastar model in the presence of RR gravity. To evaluate thermodynamic quantities in the shell, we require the value of $A(R)$, the radial component of the metric at the gravastar surface. In the Rastall–Rainbow framework, the modified Einstein field equations yield the following differential equation from the $G_t^t$ component

$$\frac{A'(r)}{rA(r)^2} - \frac{1}{r^2 A(r)} + \frac{1}{r^2} = 8\pi G\,\bar{\rho}(r). \tag{27}$$

In the shell region, the metric function $A(r)$ is known to be large due to the thin-shell approximation, i.e., $A(r) \gg 1$. To make the analysis tractable, we define

$$A^{-1}(r) \equiv h(r), \quad \text{with} \quad h(r) \ll 1. \tag{28}$$

Substituting Eq. (28) into Eq. (27) and keeping leading-order terms in $h(r)$, we obtain a simplified differential equation for $h(r)$, whose solution reads

$$h(r) = \frac{2(1-\lambda)}{\lambda}\ln r + h, \tag{29}$$

where $h$ is an integration constant determined by boundary conditions.

Consequently, the inverse relation provides the metric coefficient at the gravastar surface. Evaluating the result at $r = R$, we obtain

$$A(R) = \left[\frac{2(1-\lambda)}{\lambda}\ln R + h\right]^{-1}. \tag{30}$$

As another step, to determine the effective pressure at the shell, we utilize the temporal component of the metric Eq. (27) and substituting the thin-shell approximation Eq. (28), we obtain

$$\bar{\rho}(r) \approx \frac{1}{8\pi G r^2}. \tag{31}$$

Substituting this result into Eq. (14) and using the equation of state $p = \rho$, we express the physical energy density. Inserting this into the effective pressure relation given in Eq. (15), we find

$$\bar{p}(R) = \frac{1}{8\pi G R^2}\frac{\lambda}{(3\lambda - 2)}. \tag{32}$$

Finally, we derive the entropy of a gravastar at the shell as follows

$$S_{\text{on-shell}}^{RR} \approx \sqrt{\frac{8\pi}{G}}\,\frac{a\epsilon\,R^2}{\Sigma(\epsilon)}\sqrt{\frac{\lambda}{3\lambda-2}}\left[\frac{2(1-\lambda)}{\lambda}\ln R + h\right]^{-1/2}. \tag{33}$$

In order to make the shell entropy directly comparable to the Bekenstein-Hawking entropy, we substitute $\delta = \epsilon R$ into the final expression, where $\epsilon \ll 1$ characterizes the shell thickness relative to the gravastar radius. In the classical general relativity limit, where $\lambda \to 1$ and $\Sigma(\epsilon) \to 1$, the entropy reduces to a form proportional to $R^2$, consistent with the black hole entropy. In this limit, the parameter $h$ becomes effectively fixed by the value of $a$ and the shell thickness, $\epsilon$. This ensures that the gravastar entropy remains smaller than that of a black hole, in agreement with previous results in the main Ref. [44]. Numerically, we find that $\epsilon^2/h \approx 8.3 \times 10^{-4}$, a result which will be used later in our numerical plots and thermodynamic analysis.

To express the entropy in terms of the gravastar mass $M$, we substitute the shell radius $R$ using the interior metric given in Eq. (18). The resulting expression for the entropy reads

$$S_{\text{on-shell}}^{RR} \approx \frac{8\sqrt{2\pi}G^{3/2}aM^2}{(2\lambda-1)^2\Sigma^5(\epsilon)}\sqrt{\frac{\lambda}{3\lambda-2}}\,\frac{\epsilon}{\sqrt{h}}\left[1 - \frac{(1-\lambda)}{\lambda\,h}\ln\left(\frac{2GM}{(2\lambda-1)\Sigma^2(\epsilon)}\right)\right]. \tag{34}$$

Expanding $\Sigma(\epsilon)$ to leading order in $\gamma$ and subsequently setting $G = 1$, the entropy expression simplifies to the form

$$S_{\text{on-shell}}^{RR} \approx \frac{8\sqrt{2\pi}\,a\,M^2}{(2\lambda-1)^2}\sqrt{\frac{\lambda}{3\lambda-2}}\,\frac{\epsilon}{\sqrt{h}}\left(1 + \frac{5\gamma}{8M^2}\right)\left[1 - \frac{(1-\lambda)}{\lambda\,h}\ln\left(\frac{8M^3}{(2\lambda-1)(4M^2-\gamma)}\right)\right]. \tag{35}$$

This version highlights the dependence of the entropy on the RR parameters and demonstrates that the standard black hole entropy is recovered in the limit $\lambda = 1$ and $\gamma = 0$. A notable outcome of our analysis is that, in the RR framework, a horizonless compact object, gravastar, acquires a logarithmic correction to its entropy. The appearance of a $\ln A$ term is well established from independent approaches like CFT/Cardy analyses [56], Loop Quantum Gravity microstate counting [57], and thermal-fluctuation theory [58], supporting its near universality as the leading quantum correction. This picture is reinforced by one loop computations in the quantum entropy function for extremal black holes, where $\ln A$ arises from massless determinants and often shows matter independent coefficients [59]. Contemporary tunneling or thermal fluctuation studies of modified black holes likewise find that including the log term stabilizes the thermodynamics, typically driving the specific heat positive and the Helmholtz free energy to decrease monotonically, across Horndeski-like, magnetized Ernst-like, and gauge supergravity-like backgrounds [60–62]. Beyond its universality claim, the logarithmic correction therefore has concrete dynamical impact, and in Section 3 we will explicitly demonstrate how these corrections render the specific heat positive and smooth the free energy landscape for the RR gravastar, signaling its thermodynamic stabilization.

### A3. Exterior region

The exterior region of the gravastar, defined for $r > R + \delta$, is assumed to be a vacuum spacetime with vanishing energy density and pressure, i.e., $\rho = p = 0$, transitioning to flat spacetime at infinity. In the framework of RR gravity, we consider a modified Schwarzschild solution incorporating energy-dependent modifications through Rainbow functions. The line element in this region is given by

$$ds^2 = -\frac{1}{\Pi^2}\left(1 - \frac{2GM}{r}\right)dt^2 + \frac{1}{\Sigma^2(\varepsilon)}\left(1 - \frac{2GM}{r}\right)^{-1} dr^2 + \frac{r^2}{\Sigma^2(\varepsilon)}\, d\Omega^2, \tag{36}$$

where $M$ is the total gravitational mass of the system.

To analyze the thermodynamic behavior in the exterior region of our compact object, we employ the concept of *fiducial temperature* [63], which describes the redshifted (or blueshifted) temperature measured by a static observer located at a fixed radial distance $r$ in a gravitational field. It is worth noting that, in Ref. [64], the fiducial temperature is contrasted with the *Tolman temperature* [65], which is defined for a freely falling observer. Accordingly, a time dilation factor is applied to the RR-Hawking temperature derived on the shell in Eq. (23), yielding the redshifted temperature observed at a given location outside the gravastar. Specifically, the redshifted temperature at a radial distance $r$ is given by

$$T_{\text{fid}}^{\text{RR}}(r) = \sqrt{\frac{g_{tt}^{\text{shell}}(R)}{g_{tt}^{\text{ext}}(r)}}\, T_{\text{on-shell}}^{\text{RR}} = \sqrt{\frac{[R(1-2\lambda)\Sigma^2 - 2GM(2-3\lambda)]\, r}{R(1-2\lambda)(r - 2GM)\Sigma^2}}\, T_{\text{on-shell}}^{\text{RR}}. \tag{37}$$

Substituting the temperature on shell from Eq. (23) we obtain the following form

$$T_{\text{fid}}^{\text{RR}}(r) = \frac{(2\lambda - 1)}{4\pi R}\left(1 - \frac{\gamma}{4\, GM^2}\right)\sqrt{\frac{[R(1-2\lambda)\Sigma^2 - 2GM(2-3\lambda)]\, r}{R(1-2\lambda)(r - 2GM)\Sigma^2}}. \tag{38}$$

Obviously, by choosing $\lambda = 1$, the effects of Rastall's theory are ignored, and by setting $\gamma = 0$, the effects of Rainbow gravity are eliminated, thereby the standard Tolman temperature for an observer at a distance $r$ is recovered. Compared with GR (where $\Pi = \Sigma = 1$), the Rainbow ratio $\Sigma(\varepsilon)/\Pi(\varepsilon) \leq 1$ enters as an *energy dependent UV weight* multiplying the temperature. As $\varepsilon$ increases (equivalently, for smaller $M$), this weight decreases and bends $T(M)$; as we will show below, it produces two extrema and ultimately drives $T \to 0$ at a finite mass $M_0$. Besides, the Rastall parameter does not modify the exterior metric directly; rather, it alters the effective matter-curvature coupling, which in turn reshapes the energy density and pressure in the shell sector and readjust the shell to exterior redshift through the junction conditions.

Considering a constant $r$, we illustrate the fiducial temperature of a compact object in RR gravity versus the mass in Fig. 1. Our goal is to investigate the temperature behavior as the Rainbow and Rastall parameters vary. As we illustrate in Fig. 1, there are two critical masses where the temperatures exhibit extrema. We have specified the minimum by $(M_2, T_2)$, and the maximum by $(M_1, T_1)$, as it has been indicated on the plots. Regardless of any modification, a minimum point exists, while the maximum arises due to the presence of Rainbow gravity. Moreover, the smaller value of the Rainbow parameter weakens the UV suppression (see Eq. (2)), so the curve shifts upward and the maximum occurs at a smaller mass. Conversely, larger $\gamma$ strengthens the UV cutoff and lowers $T_{\text{max}}$. On the other hand, when the Rastall parameter is varied, as shown in the left panel of Fig. 1, the location of maximum point changes whereas the minimum mass remains nearly unaffected. The reason is that $\lambda$ adjusts the shell–to–exterior redshift through the junction conditions, whereas the remnant scale is controlled by the Rainbow factor. Specifically, an increase in the Rastall parameter leads to a higher maximum temperature occurring at a larger mass. Moreover, even in the absence of Rastall gravity, by setting $\lambda = 1$, as long as Rainbow gravity is present, the temperature still exhibits a maximum followed by a continuous decrease to approaching a remnant state with zero temperature. Conversely, when the Rastall parameter held fixed, as illustrated in the right panel of Fig. 1, increasing the Rainbow parameter strengthens the UV cutoff, which shifts the maximum toward lower temperatures and larger masses.

Moreover, a notable feature of RR gravity is the presence of an intermediate unstable phase: between the minimum and the maximum one has $dT/dM > 0$, hence $C = (\partial M/\partial T)_r < 0$. Increasing either $\lambda$ or $\gamma$ moves the two extrema closer together; beyond a critical value they merge, the positive slope interval disappears, and the unstable phase is eliminated. This behavior will be examined in detail via the heat capacity in the next subsection.

A noteworthy result in the presence of both RR gravity is the existence of a remnant, which contributes to resolving the loss information problem. If we extract the remnant mass from Eq. (38) by assuming $R = bM$, we obtain

$$M_0 = \frac{\sqrt{b\,\gamma\,(2\lambda - 1)}}{\sqrt{(2\lambda - 1)(4b - 12) + 4}}. \tag{39}$$

This relation clearly indicates the contribution of both Rainbow and Rastall parameters on the remnant mass. In the absence of Rastall gravity, the remnant mass for a black hole would simply be $\sqrt{\gamma}/2$ [23], which is clearly smaller than the remnant mass of a gravastar. However, as can be seen from Eq. (39) and Fig. 1, the inclusion of the Rastall parameter allows the remnant mass to become either smaller or larger, depending on the parameter values.

It is noteworthy that finite mass remnant with vanishing or bounded temperature appear across many quantum gravity motivated frameworks for compact objects (see for instance [66,67] and references therein). As one example among numerous works, the authors in Ref. [68] showed that implementing GUP in black hole thermodynamics changes the behavior of $T(M)$ in comparison with pure Schwarzschild case, yields a finite maximum and subsequently drives zero temperature at a nonzero mass, producing a stable remnant as a possible candidate for dark matter. Conceptually, the GUP modifies the Heisenberg relation by incorporating gravitational effects, implying a minimum length or equivalently a maximum energy(momentum), and effectively suppressing ultra-high energy modes in thermodynamic quantities. On the other hand, DSR deforms Lorentz symmetry by introducing a second invariant scale (the Planck energy/momentum), which leads to a modified dispersion relation and consequently to an energy dependent effective speed of light (Eq. (4)). More precisely, GUP acts at the level of commutators and phase space (actually with in a symplectic manifold [69]), whereas DSR operates at the level of spacetime symmetries and dispersion relations. Despite this apparent difference, both frameworks introduce a fundamental UV scale (minimum length or maximal momentum) and suppress ultra-high energy modes towards UV regularization of QFT. Authors in Ref. [70] make this connection explicit by showing how minimal length and maximal momentum realizations can be formulated consistently in Hilbert space within a DSR motivated setting to avoid divergence of energy of a free test particle (in contrast to the seminal work [71] which contains divergent free particle's energy). Consequently, the common mechanism is existence of an effective UV cutoff that reduces the contribution of the very high energy modes in thermodynamic quantities. In fact, incorporation of the DSR into the theory via existence of a maximal momentum reflects the UV regularity of the underlying QFT in the same way as the existence of a minimal length does the job in GUP framework. It is worth mentioning that these are common addresses of approaches to quantum gravity. As has been shown in [72], gravity's Rainbow can be a bridge between LQC and DSR which is an explicit example of the mentioned interconnection.

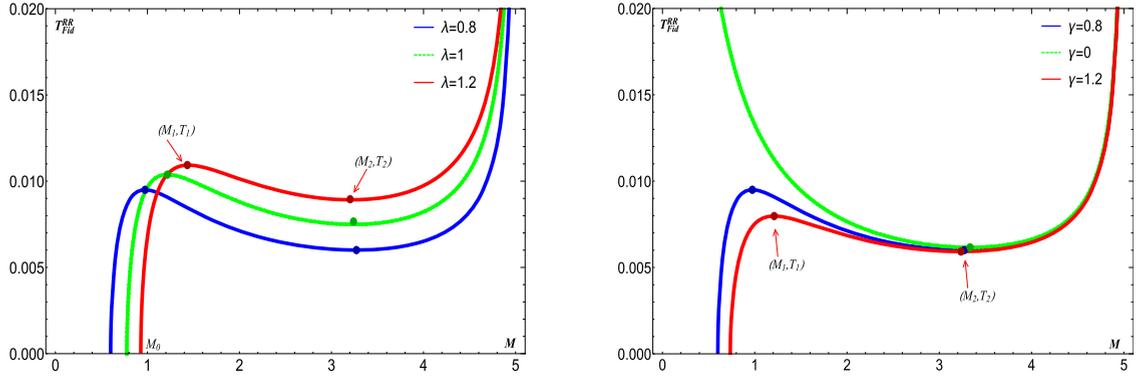

**Fig. 1.** Plots illustrate the behavior of the fiducial temperature as a function of mass. The left-hand plot is shown for different values of $\lambda = 0.8, 1, 1.2$ with fixed $\gamma = 0.8$, while the right-hand plot is shown for different values of $\gamma = 0, 0.8, 1.2$ with fixed $\lambda = 0.8$. The dashed green curves in both plots indicate the extremal condition obtained when we ignore the Rastall (left) or Rainbow (right) modifications. Other numerical settings are $G = c = 1$ and $r = 10$. (For interpretation of the references to colour in this figure legend, the reader is referred to the web version of this article.)

### 3.2. (In)stability and phase transition

#### 3.2.1. Heat capacity

In this section, we investigate additional thermodynamic properties of the compact object in the framework of RR gravity. As discussed earlier, the thermodynamic quantities such as temperature and entropy are defined locally on the intermediate thin shell of the gravastar, where the thermodynamic behavior is most significant. Assuming that the pressure outside the compact object remains fixed ($dp = 0$), and identifying the internal energy with the shell energy, the differential relation for enthalpy reduces to $dH = T\,dS$. Consequently, the heat capacity of the system can be expressed as

$$C^{RR} = \left(\frac{dH}{dT}\right)_p = \left(\frac{dE}{dT}\right) = T^{RR}_{fid} \frac{dS}{dT}. \tag{40}$$

This expression provides a valid thermodynamic relation within our model, as both $T^{RR}_{fid}$ and $S$ are derived from local shell properties. The behavior of the heat capacity allows us to probe the thermal structure of the compact object: regions with positive heat capacity correspond to thermodynamically stable phases, while negative values indicate instability. Furthermore, discontinuities or divergences in the heat capacity can signal the occurrence of a phase transition within the system. Based on these considerations, and using Eqs. (34), (37), and (40), we calculate the heat capacity of the compact object in RR gravity as follows

$$C^{RR} = 0.3\,\pi^{3/2} G^{3/2} \sqrt{\frac{\lambda}{3\lambda - 2}} (r - 2\,GM) \Big[4\gamma(1 - 5\lambda)M^2 + 32(3\lambda - 1)M^4$$
$$- 15\gamma^2(\lambda - 1) + 16(\lambda - 1)M^2\left(4M^2 - \gamma\right)\ln\left(\frac{8\,GM^3}{(2\lambda - 1)(4M^2 - \gamma)}\right)\Big]$$
$$\Big[(12\,G^2\,M^3 + 3r\gamma - GM(4Mr + 7\gamma))(1 - 2\lambda)^2\lambda\Big]^{-1}. \tag{41}$$

We derive the heat capacity as a function of the RR parameters, showing that by setting $\lambda = 1$ and $\gamma = 0$, the standard heat capacity for a compact object is recovered. As illustrated in Fig. 2, the behavior of the heat capacity reveals the existence of three distinct thermodynamic phases in RR gravity:

- $M > M_2$: Stable phase,

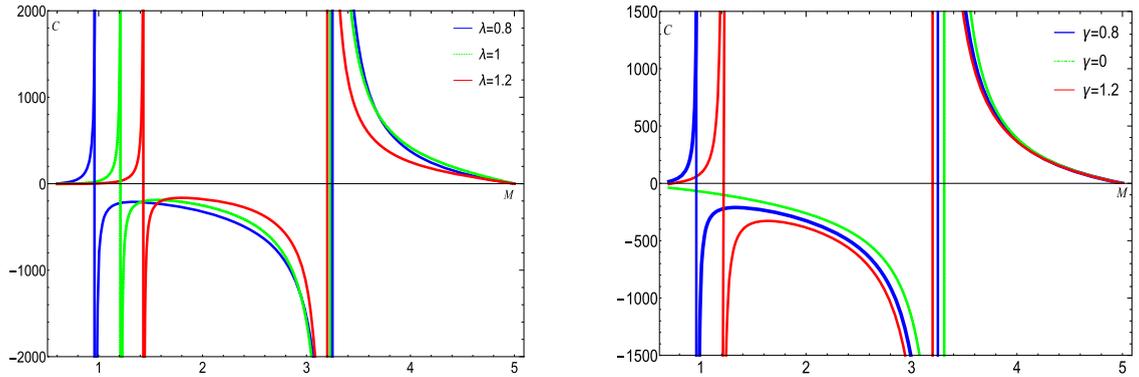

**Fig. 2.** Plots illustrate the behavior of the heat capacity as a function of mass. The left-hand plot is shown for different values of $\lambda = 0.8, 1, 1.2$ with $\gamma = 0.8$, while the right-hand plot is shown for different values of $\gamma = 0, 0.8, 1.2$ with $\lambda = 0.8$. The dashed curves indicate the pure Rainbow and pure Rastall modifications in the left and right plots, respectively.

- $M_1 < M < M_2$: Unstable phase,
- $M < M_1$: Stable phase.

In the stable phase, the compact object behaves as a conventional thermodynamic system, wherein a decrease in mass leads to a reduction in temperature and a positive heat capacity. As the mass approaches the critical value $M_2$, the system undergoes a transition into an unstable phase. Specifically, within the range $M_1 < M < M_2$, the compact object exhibits thermodynamic instability, characterized by an increase in temperature as the mass decreases, resulting in a negative heat capacity. Notably, negative heat capacity implies that, contrary to intuition, the object loses energy while its temperature rises. This paradox also arises in standard black hole thermodynamics (see the green-dashed curve in the right panel of Fig. 2, where Rainbow corrections have been neglected). The significance of RR gravity becomes apparent in this regime: the Rainbow modification prevents the system from remaining indefinitely in the unstable phase, while an increase in the Rastall parameter shortens the duration of this phase in the compact object's thermodynamic evolution.

To examine the role of the Rastall parameter, we focus on the left-hand plot of Fig. 2, where the Rainbow parameter is fixed at $\gamma = 0.8$. The plots clearly indicate that increasing the Rastall parameter shifts the onset of the unstable phase to lower mass values. According to Eq. (6), a larger value of $\lambda$ corresponds to a stronger coupling between the Ricci scalar and the energy–momentum tensor. As a result, increasing the Rastall parameter (i.e., the coupling strength) not only causes the instability to begin at smaller masses, but may also reduce the overall width of the unstable mass range. In contrast, the right-hand plot of Fig. 2 isolates the effect of the Rainbow parameter by fixing $\lambda = 0.8$. It is observed that in the absence of Rainbow modifications (i.e., pure Rastall gravity), the system does not recover stability at low masses. However, increasing $\gamma$ introduces a stabilizing effect: the system regains stability at smaller mass values, and the remnant mass becomes larger. From a physical perspective, this behavior arises because the Rainbow modification suppresses ultra-high energy modes, bending the temperature curve in such a way that the slope $dT/dM$ changes sign immediately after the maximum. This reversal flips the heat capacity and secures the stability of the small mass branch. Taken together, these results suggest complementary behavior between the two parameters. While increasing $\lambda$ reduces the upper bound of the unstable mass range, increasing $\gamma$ raises its lower bound. This interplay effectively narrows the unstable region in the mass spectrum of the compact object and accelerates the system's transition to thermodynamic stability.

### 3.2.2. Free energy

In line with the final step of our investigation, we address the Helmholtz free energy, which plays a crucial role in analyzing thermodynamic stability and phase transitions. Indeed, modern free energy analysis stems from first-order Hawking–Page phase transition of Schwarzschild-AdS black holes [73] and from York's canonical cavity formulation of Helmholtz free energy [74]. In a canonical ensemble, systems tend to minimize the Helmholtz free energy, and a transition occurs when competing branches have equal free energy. Afterward, charged AdS black holes were mapped in detail by Chamblin et al., revealing the swallowtail and a van der Waals type small/large transition in both canonical and grand canonical ensembles [75,76]. The next advances were: treating $\Lambda$ as pressure which recast black hole thermodynamics into an extended phase space with $P-V$ criticality [77], clarifying the RN-AdS Gibbs free energy swallowtail and coexistence line [78], extending the framework to charged and rotating AdS black holes and to Born–Infeld electrodynamics [79], and enriching the global phase diagram by investigating on rotating Kerr-AdS black holes and black rings [80]. In parallel, off-shell cavity studies bench marked Helmholtz free energy against hot curved space for quantum-corrected Schwarzschild [81]; massive-gravity and micro-structure results showed analogous transitions and linked coexistence lines to underlying degrees of freedom [82]. Accordingly, within Rainbow gravity, free energy calculations for Schwarzschild in an isothermal cavity showed Rainbow parameter controlled swallowtails and multiple Hawking-Page phase transitions [23]. More recently, GUP (as a quantum gravity deformation) together with the tunneling method has been used to compute corrected Helmholtz free energies for Horndeski-like [60], magnetized Ernst-like [61], and gauge-supergravity-like black holes [62]. Here, we develop the first free energy analysis of a *horizonless* compact object within RR gravity; this reveals RR controlled phase structure with small/intermediate/large branches and well-defined $F(T)$ crossing points.

**Table 1**
Description for the intersection points of free energy plots.

| Icon | Symbol | Description |
|---|---|---|
| • | $T_1$ | Intersection temperature between middle and large free energy curves |
| • | $T_2$ | Intersection temperature between small and middle free energy curves |
| ∘ | $T_c$ | Intersection temperature between free energy curves of the compact object and hot curved space |
| • | $T^*$ | Intersection temperature between small and large free energy curves |

To begin, we determine the energy of a system evolving from $m$ to $m'$ as follows

$$E^{RR} = \int_m^{m'} T^{RR} dS^{RR}, \tag{42}$$

Using Eqs. (34) and (38), we derive the energy observable at a distance $r$ from a compact object in RR gravity as follows

$$E^{RR} = \frac{2\sqrt{2}\, G^{1/2} a}{3\sqrt{\pi}\,\lambda\,(2\lambda-1)^2 \Sigma^3} \sqrt{\frac{\lambda}{3\lambda-2}} \frac{\epsilon}{h^{3/2}} \sqrt{\frac{\left[3(1-2\lambda)\Sigma^2 - 2G(2-3\lambda)\right]r}{3(1-2\lambda)(r-2GM)\Sigma^2}} \tag{43}$$

$$\left[(2\,GM - r)\left(2(\lambda-1)\ln\left(\frac{2\,GM}{(2\lambda-1)\Sigma^2}\right) + (2h-3)\lambda + 3\right)\right.$$

$$\left. + 4(\lambda-1)\sqrt{r(2\,GM-r)}\,\tan^{-1}\left(\sqrt{\frac{2\,GM}{r}-1}\right)\right]\Bigg|_{M_0}^{M}.$$

It makes sense to choose the lower mass limit as $m = M_0$ from Eq. (39), and the upper limit as $m' = M$. Furthermore, by neglecting the Rastall and Rainbow modifications, i.e., setting $\lambda = 1$ and $\gamma = 0$, one can recover the standard internal energy of a canonical ensemble. Subsequently, additional insights into stability and phase transitions can be obtained using the Helmholtz free energy. In the context of the canonical ensemble, the Helmholtz free energy of spacetime is given by

$$F^{RR} = E^{RR} - T^{RR}_{fid}\, S^{RR}. \tag{44}$$

Considering Eqs. (34), (38), and (43), we compute the free energy of a compact object in RR gravity numerically for fixed values of $\gamma$ and $\lambda$. This formulation serves as a powerful tool to investigate the thermodynamic behavior of a system composed of a compact object in equilibrium with a heat bath at fixed fiducial temperature, $T_{fid}$. In particular, Eq. (44) expresses the free energy as a function of mass as well as the Rainbow and Rastall parameters. For the purpose of analysis, it is useful to invert the temperature-mass relation and express the free energy as a function of temperature. To do this, we extract three distinct mass branches from Eq. (38), corresponding to three possible configurations for a given temperature:

- $M < M_1$: **Small branch** – red curves in Fig. 3
- $M_1 < M < M_2$: **Middle branch** – green curves in Fig. 3
- $M > M_2$: **Large branch** – blue curves in Fig. 3

As we demonstrate, these labels – small, middle, and large – not only reflect the mass and size of the compact object but also correspond to distinct thermodynamic properties, particularly temperature and stability. Consequently, we obtain three free energy expressions as functions of temperature, which are illustrated as the coexistence curves in Fig. 3.

As can clearly be observed, the branches of the free energy curves tend to merge or converge at certain critical temperatures, as detailed in Table 1. Before analyzing the behavior of these plots, it is important to clarify the role of the horizontal dashed line, which represents the free energy of spacetime itself, commonly referred to as the free energy of hot curved space. This corresponds to a background configuration in which a massive object and hypothetical thermal radiation are in equilibrium. Although in certain quantum-deformed gravity theories this background free energy may be nonzero (as discussed in Refs. [82,83]), in our case, evaluating Eq. (44) with the rest mass yields $F_0 = 0$. This reference value is essential when comparing a compact object's free energy to that of a radiation-dominated background. As mentioned before, phase transitions and thermodynamic stability can be examined by identifying the state with the lowest free energy. Such states represent configurations that the system naturally settles into, as they possess the lowest free energy under the conditions defined by the ensemble.

To explore the effects of RR gravity on the stability and phase transitions of a compact object, we present plots (a) through (d) corresponding to specific values of the Rastall and Rainbow parameters. These plots can be interpreted as follows:

The green branches in all plots correspond to an intermediate state whose free energy lies between those of the large (blue) and small (red) configurations. This state exhibits negative heat capacity and clearly represents a thermodynamically unstable phase that facilitates the transition between the small and large states. From zero temperature, corresponding to the remnant mass of the gravastar, up to the transition point $T_1$, the small gravastar configuration is thermodynamically preferred. Its lower free energy compared to the background spacetime and its positive heat capacity both indicate a unique feature: a stable remnant phase induced by the presence of Rainbow and Rastall modifications. In comparison, while comprehensive studies of free energy exist in Refs. [79,80,82], and even GUP based works in Refs. [60–62] despite computing thermodynamic corrections and obtaining the remnant with finite temperature, none of these analyses benchmark the small black hole branch against hot curved space and this global stability question is not addressed. Although such a stable small black hole branch has not been observed for uncharged cases, a notable exception arises in the canonical (fixed charge) analysis of RN-AdS, where a stable small branch is identified from free-energy diagrams [75,76]. The emergence of this stable branch in the RN–AdS framework bears a striking conceptual resemblance to the stabilization of the small RR gravastar, hinting at a deeper correspondence between the two scenarios. Against this backdrop, the free energetic stability we find for the RR gravastar is especially striking when contrasted with Rainbow gravity black holes, where the small branch can be locally stable yet not globally preferred. In Ref. [23], for such a Rainbow modified black hole, it was shown that the system exhibits more subtle thermodynamic behavior, making it difficult to definitively assess its stability. This ambiguity stems from an apparent mismatch between the heat capacity and free energy results in the corresponding regime. Indeed, while the small black hole has a positive heat capacity, indicating local stability, its higher free energy compared to the hot curved background renders it thermodynamically unfavorable. In contrast, in our analysis of the gravastar within RR gravity, or even under a pure Rainbow modification as shown in panel (b), the small configuration is found to be both locally and globally stable, exhibiting a positive heat capacity along with a free energy lower than that of the hot curved space. This key difference highlights a thermodynamic advantage of the gravastar over its black hole counterpart and serves as one of the notable contributions of this work. Physically, the stable small branch results from the interplay of the Rainbow modification, the Rastall coupling, and the gravastar structure itself. The Rainbow sector provides an energy dependent UV suppression that bends $T(M)$, thus guaranteeing the existence of a remnant. Most crucially, the gravastar differs from a black hole in that the absence of a horizon confines the entropy budget to the thin shell, with the de Sitter core contributing negligibly. Consequently, the gravastar's layered structure and the shell's pressure-density profile drive the internal energy below that of hot curved space; together with the small shell entropy that keeps the $TS$ term subdominant, these effects make $F$ closely track $E$ and fall below the baseline. This mechanism secures the global preference of the small RR gravastar branch, unlike in black holes, where the combination of a large positive internal energy and a dominant $TS$ term keeps the free energy above the hot curved space baseline, leaving the small BH branch globally disfavored.

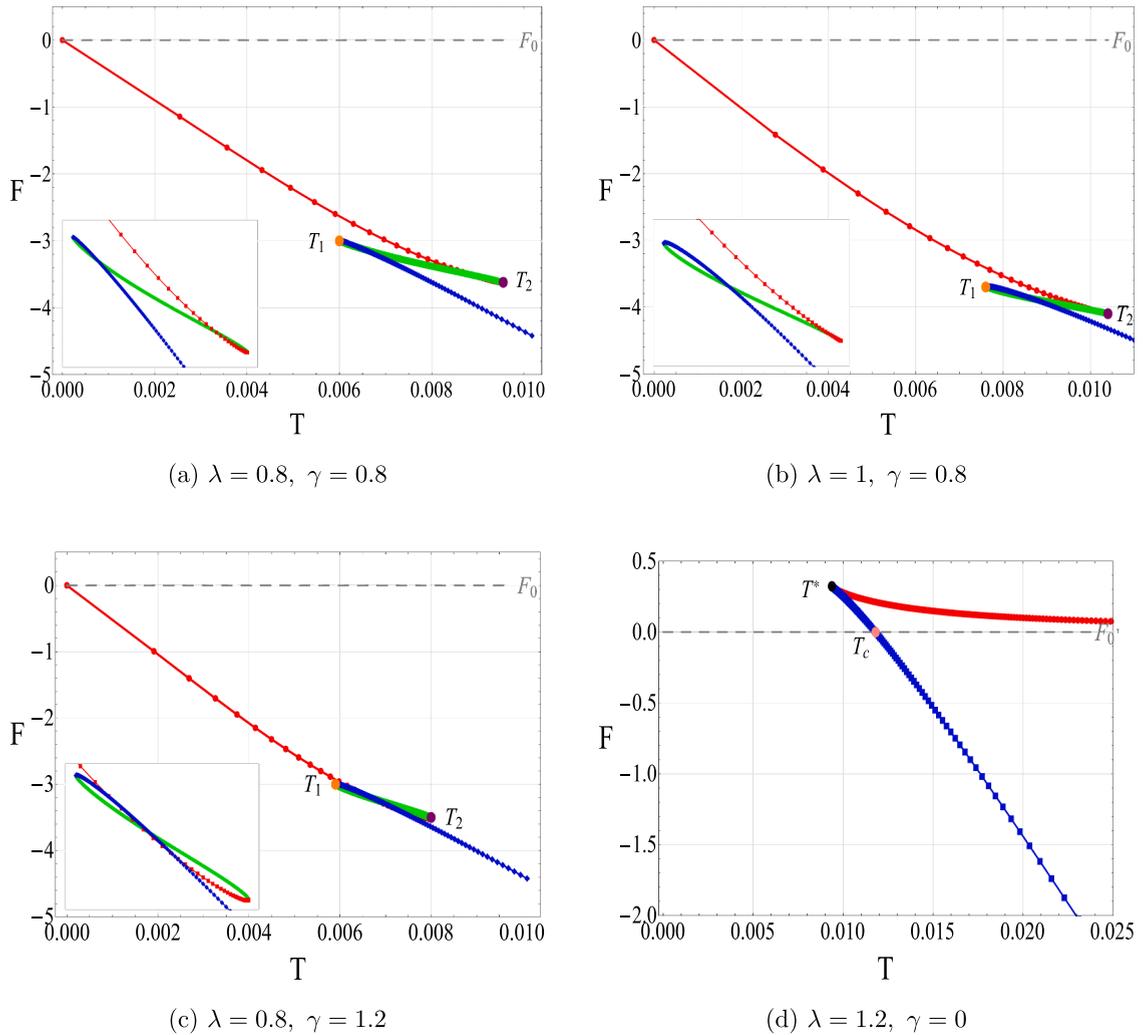

**Fig. 3.** Plots show the Helmholtz free energy as a function of temperature, which are essential for identifying the thermodynamically preferred states. In the first row, the Rainbow parameter is fixed at $\gamma = 0.8$ and the Rastall parameter increases from $\lambda = 0.8$ in panel (a) to $\lambda = 1.0$ in panel (b). In panel (c), the Rastall parameter is fixed at $\lambda = 0.8$ while the Rainbow parameter is increased to $\gamma = 1.2$. Panel (d) represents the pure Rastall case with $\gamma = 0$, i.e., no Rainbow correction. In all plots, the red, green, blue, and gray dashed curves correspond to the free energy of the small, middle, and large compact-object states, and of the hot curved space, respectively. The dots indicate the critical temperatures defined in Table 1. The numerical parameters are $G = 1$ and $r = 10$. (For interpretation of the references to colour in this figure legend, the reader is referred to the web version of this article.)

As illustrated in Fig. 3, the presence of both RR modifications results in a well-defined pattern of thermodynamic preferences: the small gravastar configuration is favored at temperatures below $T_1$, while the large configuration becomes preferred above $T_2$. In addition to shifting the critical temperatures $T_1$ and $T_2$, varying the strength of the two modification theories also alters the nature of the phase transition. In panel (a), the energy curves suggest that the small state tends to decay into the middle and large configuration within the temperature interval $T_1 < T < T_2$, passing through a thermodynamically unstable intermediate state. In panel (b), increasing the Rastall parameter $\lambda$ shifts both critical temperatures upward and reduces the width of the phase-transition interval. Physically, as the matter-curvature coupling strengthens, the shell's energy-pressure distribution is reshaped in a way that diminishes the unstable phase and narrows the unstable interval. Conversely, in panel (c), increasing the Rainbow parameter $\gamma$, which enhances the energy dependence of the spacetime geometry and thus amplifies the role of quantum gravitational effects (UV cutoff), leads to a noticeable shift of $T_2$ toward lower temperatures. Moreover, the phase transition process becomes more intricate, involving a broader and more complex transition between the small and large configurations.

If the quantum gravitational effects of Rainbow are turned off by setting $\gamma = 0$, as shown in panel (d), two distinct temperatures emerge: the critical temperature $T_c$ and $T^*$. Notably, $T^*$ can be interpreted as the temperature of a first-order Hawking-Page phase transition from a small to a large black hole in AdS spacetime [73]. In the intermediate temperature range $T^* < T < T_c$, both the small and large configurations tend to decay into the hot curved space state. At temperatures above $T_c$, the system once again thermodynamically favors the large compact object configuration. Within this pure Rastall compact object, the small configuration, characterized by a negative heat capacity, cannot form as a stable state because the hot curved space with hypothetical radiation is thermodynamically preferred. It is worth noting that in the pure Rastall limit our qualitative stability picture is consistent with Ref. [13], where the exterior region mimics a Schwarzschild-de Sitter spacetime and stability follows from surface redshift and entropy maximization criteria. Therefore, the significance of the Rainbow modification lies in its ability to predict a stable remnant for the compact object by UV suppression, making the resulting thermodynamics more realistic compared to scenarios where only a radiation-dominated background remains.

**Future outlook.** The present analysis can be extended in several directions. A useful assessment of robustness is to extend our free energy and stability analysis for alternative Rainbow profiles beyond the ansatz used in Eq. (2) e.g., the Amelino–Camelia type, GRB-motivated, and Magueijo–Smolin invariant speed families in Ref. [41] for a side by side comparison in black hole settings. Another natural extension is to embed the RR gravastar in dark energy cosmological background like quintom and compare its thermodynamics with our result.

## 4. Summary and conclusion

In this paper, we probe the thermodynamic features of a compact object in the RR modification content. On one hand, the Rainbow modifications are introduced by adopting the DSR; on the other hand, the Rastall modifications are incorporated by accepting the non-conservation of the energy–momentum tensor. The Rainbow parameter is introduced through the MDR in Eq. (1), while the Rastall parameter in Eq. (6) represents the strength of the coupling between the energy-momentum tensor and curvature. Employing the spherically symmetric metric of Ref. [30], which incorporates both RR modifications, together with the gravastar construction originally proposed in Ref. [43], we set up a three-region configuration within RR gravity for our analysis.

After determining the temperature and entropy of the intermediate shell, we focused on calculating the temperature at a specific distance from the compact object, $T_{fid}$, as well as its heat capacity and free energy. The fiducial temperature demonstrates the presence of two extrema, maximum $(M_1, T_1)$ and minimum $(M_2, T_2)$, as we pointed out in Fig. 1. Also, the plot confirms the existence of a stable mass remnant, $M_0$, with zero temperature. On one hand, an increase in the Rastall parameter shifts the maximum temperature to higher values and pushes it to occur at larger masses, since stronger matter-curvature coupling enhances the effective redshift, while the minimum mass remains nearly unchanged. On the other hand, increasing the Rainbow parameter leaves the minimum point almost unaffected, whereas the maximum shifts to larger masses and lower temperatures, because stronger UV cutoff shifts the onset of suppression to larger masses and lower temperatures. Moreover, increasing either the Rastall or Rainbow parameter causes the extrema of the fiducial temperature curve to converge, indicating the presence of an upper bound beyond which the unstable intermediate phase disappears.

In the next step, we divided spacetime into three stable or unstable phases by calculating and plotting the heat capacity in Fig. 2. We deduced that while the pure Rastall cannot prevent the final unstable phase, variation in both parameters can change the structure of stabilities. In physical terms, the Rainbow modification suppresses the contribution of ultra-high energy modes. This suppression bends the temperature curve in such a way that the slope $dT/dM$ reverses immediately after the maximum, causing the heat capacity to flip sign and thereby stabilizing the small mass branch through the emergence of a remnant. Whereas the Rastall coupling, through its modification of the matter-curvature interaction, reshapes the shell's energy-pressure distribution and thereby narrows the unstable interval.

In the following, for a deeper investigation into the stability stages, we computed and plotted the Helmholtz free energy of the system in Fig. 3. To illustrate the free energy as a function of temperature, we first determined the mass function for each region, categorized as small, middle, and large. Initially, the free energy plots also confirm the existence of an unstable phase during the middle stage of the compact object's life cycle. Remarkably, a stable small configuration emerges at low temperatures, highlighting the distinctive thermodynamic characteristics of gravastars in RR gravity compared to black holes. In physical terms, the stable small branch results from the interplay of the Rainbow modification, which suppresses ultra-high energy modes, the Rastall coupling, which adjusts the matter-curvature balance, and the gravastar structure itself, which confines the entropy to the thin shell. Parametrically, increasing the Rastall parameter raises both critical temperatures and shortens the range of the phase transition. It also makes the free energy of the small state decrease more rapidly with temperature. In contrast, increasing the Rainbow parameter by enhancing the UV suppression significantly impacts on the two critical temperatures, bringing the energies of the middle and large states closer to the small state's free energy. This behavior results in a richer structure of phase transitions, including the emergence of a first-order Hawking-Page-like transition.

**Data availability**

No data was used for the research described in the article.

**Declaration of competing interest**

The authors declare that they have no known competing financial interests or personal relationships that could have appeared to influence the work reported in this paper.

**Acknowledgment**

We are deeply grateful to Kourosh Nozari for his valuable intellectual contributions and constructive feedback, which significantly enriched the development of this work.

**References**

[1] P. Rastall, Generalization of the Einstein theory, Phys. Rev. D 6 (1972) 3357. https://doi.org/10.1103/PhysRevD.6.3357
[2] A.S. Al-Rawaf, M.O. Taha, Cosmology of general relativity without energy-momentum conservation, Gen. Relativ. Grav. 28 (1996) 935. https://doi.org/10.1007/BF02113090
[3] M. Capone, V.F. Cardone, M.L. Ruggiero, The possibility of an accelerating cosmology in Rastall's theory, J. Phys. Conf. Ser. 222 (2010) 12012. https://doi.org/10.1088/1742-6596/222/1/012012
[4] C.E.M. Batista, M.H. Daouda, J.C. Fabris, et al., Rastall cosmology and the $\Lambda$ CDM model, Phys. Rev. D 85 (2012) 84008. https://doi.org/10.1103/PhysRevD.85.084008
[5] J.C. Fabris, M.H. Daouda, O.F. Piattella, Note on the evolution of the gravitational potential in Rastall scalar field theories, Phys. Lett. B 711 (2012) 232. https://doi.org/10.1016/j.physletb.2012.04.020
[6] F. Darabi, K. Atazadeh, Y. Heydarzade, Einstein static universe in the rastall theory of gravity, Eur. Phys. J. Plus 133 (2018) 249. https://doi.org/10.1140/epjp/i2018-12083-1
[7] H. Moradpour, Y. Heydarzade, F. Darabi, et al., A generalization to the rastall theory and cosmic eras, Eur. Phys. J. C 77 (2017) 259. https://doi.org/10.1140/epjc/s10052-017-4811-z
[8] J. Astorga-Moreno, K. Jacobo, S. Arteaga, et al., $\Lambda$ CDM-Rastall Cosmology revisited: constraints from a recent Quasars datasample, Class. Quant. Grav. 41 (2024) 65003. https://doi.org/10.1088/1361-6382/ad1fca
[9] A.M. Oliveira, H.E.S. Velten, J.C. Fabris, et al., Neutron stars in Rastall gravity, Phys. Rev. D 92 (2015) 44020. https://doi.org/10.1103/PhysRevD.92.044020
[10] K.A. Bronnikov, J.C. Fabris, O.F. Piattella, et al., Static, spherically symmetric solutions with a scalar field in rastall gravity, Gen. Relativ. Grav. 48 (2016) 162. https://doi.org/10.1007/s10714-016-2152-0
[11] Y. Heydarzade, F. Darabi, Black hole solutions surrounded by perfect fluid in Rastall theory, Phys. Lett. B 771 (2017) 365. https://doi.org/10.1016/j.physletb.2017.05.064
[12] G. Abbas, M.R. Shahzad, A new model of quintessence compact stars in the Rastall theory of gravity, Eur. Phys. J. A 54 (2018) 211. https://doi.org/10.1140/epja/i2018-12642-y
[13] S. Ghosh, S. Dey, A. Das, et al., Study of gravastars in Rastall gravity, JCAP 07 (2021) 4. https://doi.org/10.1088/1475-7516/2021/07/004
[14] H. Moradpour, I.G. Salako, Thermodynamic analysis of the static spherically symmetric field equations in Rastall theory, 2016, p. 3492796. https://doi.org/10.1155/2016/3492796
[15] I.P. Lobo, H. Moradpour, J.P.M. Graça, et al., Thermodynamics of black holes in Rastall gravity, Int. J. Mod. Phys. D 27 (2018) 1850069. https://doi.org/10.1142/S0218271818500694
[16] S. Soroushfar, M. Afrooz, Analytical solutions of the geodesic equation in the spacetime of a black hole surrounded by perfect fluid in Rastall theory, Indian J. Phys. 96 (2022) 593. https://doi.org/10.1007/s12648-020-01971-5
[17] R. Ali, M. Asgher, M.F. Malik, Gravitational analysis of neutral regular black hole in Rastall gravity, Mod. Phys. Lett. A 35 (2020) 2050225. https://doi.org/10.1142/S0217732320502259
[18] J. Magueijo, L. Smolin, Lorentz invariance with an invariant energy scale, Phys. Rev. Lett. 88 (2002) 190403. https://doi.org/10.1103/PhysRevLett.88.190403
[19] G. Amelino-Camelia, Doubly-Special Relativity: Facts, Myths and Some Key Open Issues, 2, 2010. https://doi.org/10.3390/sym2010230
[20] J. Magueijo, L. Smolin, Gravity'S rainbow, Class. Quant. Grav 21 (2004) 1725. https://doi.org/10.1088/0264-9381/21/7/001
[21] Y. Ling, X. Li, H.B. Zhang, Thermodynamics of modified black holes from gravity's rainbow, Mod. Phys. Lett. A 22 (2007) 2749. https://doi.org/10.1142/S0217732307022931

[22] C.Z. Liu, J.Y. Zhu, Hawking radiation and black hole entropy in a gravity's rainbow, Gen. Relativ. Grav. 40 (2008) 1899. https://doi.org/10.1007/s10714-008-0607-7.
[23] Z.W. Feng, S.Z. Yang, Thermodynamic phase transition of a black hole in rainbow gravity, Phys. Lett. B 772 (2017) 737. https://doi.org/10.1016/j.physletb.2017.07.057.
[24] B.E. Panah, Effects of energy dependent spacetime on geometrical thermodynamics and heat engine of black holes: gravity's rainbow, Phys. Lett. B 787 (2018) 45. https://doi.org/10.1016/j.physletb.2018.10.042.
[25] A.F. Ali, M. Faizal, M.M. Khalil, Remnant for all black objects due to gravity's rainbow, Nucl. Phys. B 894 (2015) 341. https://doi.org/10.1016/j.nuclphysb.2015.03.014.
[26] C. Leiva, J. Saavedra, J. Villanueva, The geodesic structure of the Schwarzschild black holes in gravity's rainbow 24 (2009) 1443. https://doi.org/10.1142/S0217732309029983.
[27] A.F. Ali, M. Faizal, B. Majumder, Absence of an effective horizon for black holes in gravity's rainbow, Europhys. Lett. 109 (2015) 20001. https://doi.org/10.1209/0295-5075/109/20001.
[28] Y. Gim, W. Kim, Black hole complementarity in gravity's rainbow, JCAP 05 (2015) 2. https://doi.org/10.1088/1475-7516/2015/05/002.
[29] A.B. Tudeshki, G.H. Bordbar, B.E. Panah, Effect of rainbow function on the structural properties of dark energy star, Phys. Lett. B 848 (2024) 138333. https://doi.org/10.1016/j.physletb.2023.138333.
[30] C.E. Mota, L.C.N. Santos, G. Grams, Combined Rastall and rainbow theories of gravity with applications to neutron stars, Phys. Rev. D 100 (2019) 24043. Grams et al. https://doi.org/10.1103/PhysRevD.100.024043.
[31] C.E. Mota, L.C.N. Santos, F.M.D. Silva, et al., Anisotropic compact stars in Rastall-rainbow gravity, Class. Quant. Grav. 39 (2022) 85008. https://doi.org/10.1088/1361-6382/ac5a13.
[32] R.C. Tolman, Static solutions of Einstein's field equations for spheres of fluid, Phys. Rev. 55 (1939) 364. https://doi.org/10.1103/PhysRev.55.364.
[33] J.R. Oppenheimer, G.M. Volkoff, On massive neutron cores, Phys. Rev. 55 (1939). https://doi.org/10.1103/PhysRev.55.374.
[34] U. Debnath, Charged gravastars in Rastall-rainbow gravity, Eur. Phys. J. Plus 136 (2021) 640. https://doi.org/10.1140/epjp/s13360-021-01460-6.
[35] K.P. Das, S. Maity, P. Saha, et al., Charged anisotropic strange star in Rastall-Rainbow gravity, Mod. Phys. Lett. A 37 (2022) 2250201. https://doi.org/10.1142/S0217732322502017.
[36] J. Li, B. Yang, W. Lin, Massive white dwarfs in Rastall-Rainbow gravity, JCAP 04 (2024) 81. https://doi.org/10.1088/1475-7516/2024/04/081.
[37] O.P. Jyothilakshmi, L.J. Naik, V. Sreekanth, Bose-Einstein Condensate Stars in Combined Rastall-Rainbow Gravity, 2023. https://doi.org/10.48550/arXiv.2311.13813.
[38] T. Tangphati, D.J. Gogoi, A. Pradhan, et al., Investigating stable quark stars in Rastall-Rainbow gravity and their compatibility with gravitational wave observations, J. High Energy Astrophys. 42 (2024) 12. https://doi.org/10.1016/j.jheap.2024.02.006.
[39] G. Amelino-Camelia, J.R. Ellis, N.E. Mavromatos, et al., Distance measurement and wave dispersion in a Liouville string approach to quantum gravity, Int. J. Mod. Phys. A 12 (1997) 607. https://doi.org/10.1142/S0217751X97000566.
[40] J. Magueijo, L. Smolin, Generalized Lorentz invariance with an invariant energy scale, Phys. Rev. D 67 (2003) 44017. https://doi.org/10.1103/PhysRevD.67.044017.
[41] A.F. Ali, M.M. Khalil, A proposal for testing gravity's rainbow, EPL 110 (2015). https://doi.org/10.1209/0295-5075/110/20009.
[42] M. Khodadi, K. Nozari, A consistency check for the free scalar field theory realization of the doubly special relativity, Commun. Theor. Phys. 71 (2019) 677. https://doi.org/10.1088/0253-6102/71/6/677.
[43] P.O. Mazur, E. Mottola, Gravitational vacuum condensate stars, Proc. Natl. Acad. Sci. U.S.A. 101 (2004) 9545. https://doi.org/10.1073/pnas.0402717101.
[44] P.O. Mazur, E. Mottola, Gravitational Condensate Stars: An Alternative to Black Holes, 2023. https://doi.org/10.3390/universe9020088.
[45] N. Sakai, H. Saida, T. Tamaki, Gravastar shadows, Phys. Rev. D 90 (2014) 104013. https://doi.org/10.1103/PhysRevD.90.104013.
[46] C. Chirenti, L. Rezzolla, Did GW150914 produce a rotating gravastar?, Phys. Rev. D 94 (2016) 84016. https://doi.org/10.1103/PhysRevD.94.084016.
[47] T. Kubo, N. Sakai, Gravitational lensing by gravastars, Phys. Rev. D 93 (2016) 84051. https://doi.org/10.1103/PhysRevD.93.084051.
[48] K.I. Nakao, K. Okabayashi, T. Harada, Radiative gravastar with Gibbons-Hawking temperature, Phys. Rev. D 105 (2022) 84017. https://doi.org/10.1103/PhysRevD.105.084017.
[49] K.I. Nakao, K. Okabayashi, T. Harada, Radiative gravastar with thermal spectrum; Sudden vacuum condensation without gravitational collapse, Phys. Rev. D 106 (2022) 44016. https://doi.org/10.1103/PhysRevD.106.044016.
[50] N. Uchikata, S. Yoshida, P. Pani, Tidal deformability and I-Love-Q relations for gravastars with polytropic thin shells, Phys. Rev. D 94 (2016) 64015. https://doi.org/10.1103/PhysRevD.94.064015.
[51] K. Akiyama, et al., Event horizon telescope collaboration, first M87 event horizon telescope results. i. the shadow of the supermassive black hole, Astrophys. J. Lett. 875 (2019) 1. https://doi.org/10.3847/2041-8213/ab0ec7.
[52] Y. Mizuno, Z. Younsi, C.M. Fromm, et al., The current ability to test theories of gravity with black hole shadows, Nat. Astron. 2 (2018) 585. https://doi.org/10.1038/s41550-018-0449-5.
[53] A. Debenedictis, Some Singular Spacetimes and their Possible Alternatives, 7, 2024. https://doi.org/10.3390/particles7040054.
[54] K.P. Das, U. Debnath, S. Ray, Thin-shell gravastars in the effect of graviton mass: a study on linearized stability and dynamics, Phys. Dark Univ. 46 (2024) 101691. https://doi.org/10.1016/j.dark.2024.101691.
[55] G. Mustafa, F. Javed, S.K. Maurya, et al., Imprints of dark energy models on structural properties of charged gravastars in extended teleparallel gravity, Phys. Dark Univ. 46 (2024) 101574. https://doi.org/10.1016/j.dark.2024.101574.
[56] S. Carlip, Logarithmic corrections to black hole entropy from the Cardy formula, Class. Quant. Grav. 17 (2000) 4175. https://doi.org/10.1088/0264-9381/17/20/302.
[57] R.K. Kaul, P. Majumdar, Logarithmic correction to the Bekenstein-Hawking entropy, Phys. Rev. Lett. 84 (2000) 5255. https://doi.org/10.1103/PhysRevLett.84.5255.
[58] S. Das, P. Majumdar, R.K. Bhaduri, General logarithmic corrections to black hole entropy, Class. Quant. Grav. 19 (2002) 2355. https://doi.org/10.1088/0264-9381/19/9/302.
[59] A. Sen, Quantum entropy function, logarithmic corrections to black hole entropy and precision counting of microstates, JHEP 04 (2013) 156. https://doi.org/10.1007/JHEP04(2013)156.
[60] R. Ali, Z. Akhtar, R. Babar, et al., Logarithm corrections and thermodynamics for Horndeski gravity like black holes, New Astron. 100 (2023) 101976. https://doi.org/10.1016/j.newast.2022.101976.
[61] R. Ali, Z. Akhtar, K. Bamba, et al., Tunneling and thermodynamics evolution of the magnetised Ernst-like black hole, Gen. Relativ. Grav. 55 (2023) 28. https://doi.org/10.1007/s10714-023-03080-0.
[62] R. Ali, M. Asgher, P.K. Sahoo, Study of tunneling radiation and thermal fluctuations of a gauge super gravity like black hole, Ann. Phys. 452 (2023) 169283. https://doi.org/10.1016/j.aop.2023.169283.
[63] S. Weinberg, Gravitation and Cosmology: Principles and Applications of the General Theory of Relativity, Wiley, New York, New York, 1972.
[64] Y. Gim, W. Kim, A quantal Tolman temperature, Eur. Phys. J. C 75 (2015) 549. https://doi.org/10.1140/epjc/s10052-015-3765-2.
[65] R.C. Tolman, P. Ehrenfest, Temperature equilibrium in a static gravitational field, Phys. Rev. 36 (1930) 1791. https://doi.org/10.1103/PhysRev.36.1791.
[66] K. Nozari, S. Islamzadeh, Tunneling of massive and charged particles from noncommutative Reissner-Nordström black hole, Astrophys. Space Sci. 347 (2013) 299. https://doi.org/10.1007/s10509-013-1532-0.
[67] S. Eslamzadeh, J.T. Firouzjaee, K. Nozari, Radiation from Einstein-Gauss-Bonnet de Sitter black hole via tunneling process, Eur. Phys. J. C 82 (2022) 75. https://doi.org/10.1140/epjc/s10052-022-09992-6.
[68] K. Nozari, H. Mehdipour, Gravitational uncertainty and black hole remnants, Mod. Phys. Lett. A 20 (2005) 2937. https://doi.org/10.1142/S0217732305018050.

[69] K. Nozari, M.A. Gorji, V. Hosseinzadeh, Natural cutoffs via compact symplectic manifolds, Class. Quant. Grav 33 (2016) 25009. https://doi.org/10.1088/0264-9381/33/2/025009.
[70] K. Nozari, A. Etemadi, Minimal length, maximal momentum and Hilbert space representation of quantum mechanics, Phys. Rev. D 85 (2012) 104029. https://doi.org/10.1103/PhysRevD.85.104029.
[71] A. Kempf, G. Mangano, R.B. Mann, Hilbert space representation of the minimal length uncertainty relation, Phys. Rev. D 52 (1995) 1108. https://doi.org/10.1103/PhysRevD.52.1108.
[72] M.A. Gorji, K. Nozari, B. Vakili, Gravity's rainbow: a bridge between LQC and DSR, Phys. Lett. B 765 (2017) 113. https://doi.org/10.1016/j.physletb.2016.12.023.
[73] S.W. Hawking, D.N. Page, Thermodynamics of black holes in anti-de Sitter space, Commun. Math. Phys. 87 (1983) 577. https://doi.org/10.1007/BF01208266.
[74] J.W.Y. Jr, W. James, Black hole thermodynamics and the euclidean Einstein action, Phys. Rev. D 33 (1986) 2092. https://doi.org/10.1103/PhysRevD.33.2092.
[75] A. Chamblin, R. Emparan, C.V. Johnson, et al., Charged AdS black holes and catastrophic holography, Phys. Rev. D 60 (1999) 64018. https://doi.org/10.1103/PhysRevD.60.064018.
[76] A. Chamblin, R. Emparan, C.V. Johnson, et al., Holography, thermodynamics and fluctuations of charged AdS black holes, Phys. Rev. D 60 (1999) 104026. https://doi.org/10.1103/PhysRevD.60.104026.
[77] D. Kubizňák, R.B. Mann, P-V criticality of charged AdS black holes, JHEP 07 (2012) 33. https://doi.org/10.1007/JHEP07(2012)033.
[78] S.W. Wei, Y.X. Liu, Insight into the microscopic structure of an AdS black hole from a thermodynamical phase transition, Phys. Rev. Lett. 115 (2015) 111302. https://doi.org/10.1103/PhysRevLett.115.111302.
[79] S. Gunasekaran, R.B. Mann, D. Kubizňák, Extended phase space thermodynamics for charged and rotating black holes and Born-Infeld vacuum polarization, 11, 2012, p. 110. https://doi.org/10.1007/JHEP11(2012)110.
[80] N. Altamirano, D. Kubizňák, R.B. Mann, et al., Thermodynamics of Rotating Black Holes and Black Rings: Phase Transitions and Thermodynamic Volume, 2, 2014. https://doi.org/10.3390/galaxies2010089.
[81] W. Kim, Y. Kim, Phase transition of quantum corrected Schwarzschild black hole, Phys. Lett. B 718 (2012) 687. https://doi.org/10.1016/j.physletb.2012.11.017.
[82] R.G. Cai, Y.P. Hu, Q.Y. Pan, et al., Thermodynamics of black holes in massive gravity, Phys. Rev. D 91 (2015) 24032. https://doi.org/10.1103/PhysRevD.91.024032.
[83] D.I. Kazakov, S.N. Solodukhin, On quantum deformation of the Schwarzschild solution, Nucl. Phys. B 429 (1994) 153. https://doi.org/10.1016/S0550-3213(94)80045-6.